\documentclass[aps,
prd,
english,
nofootinbib,
reprint]{revtex4-1}

\usepackage{amsfonts,amsmath,amssymb}
\usepackage{graphicx}
\usepackage[utf8]{inputenc}
\usepackage{hyperref}
\usepackage{babel}
\usepackage{enumitem}

\newcommand\e{{\rm e}}

\newcommand\be{\begin{equation}}
\newcommand\ee{\end{equation}}
\newcommand\bea{\begin{eqnarray}}
\newcommand\eea{\end{eqnarray}}

\begin{document}

\def\rhoo{\rho_{_0}\!} 
\def\rhooo{\rho_{_{0,0}}\!} 

\begin{flushright}
\phantom{
{\tt arXiv:2006.$\_\_\_\_$}
}
\end{flushright}

{\flushleft\vskip-1.4cm\vbox{\includegraphics[width=1.15in]{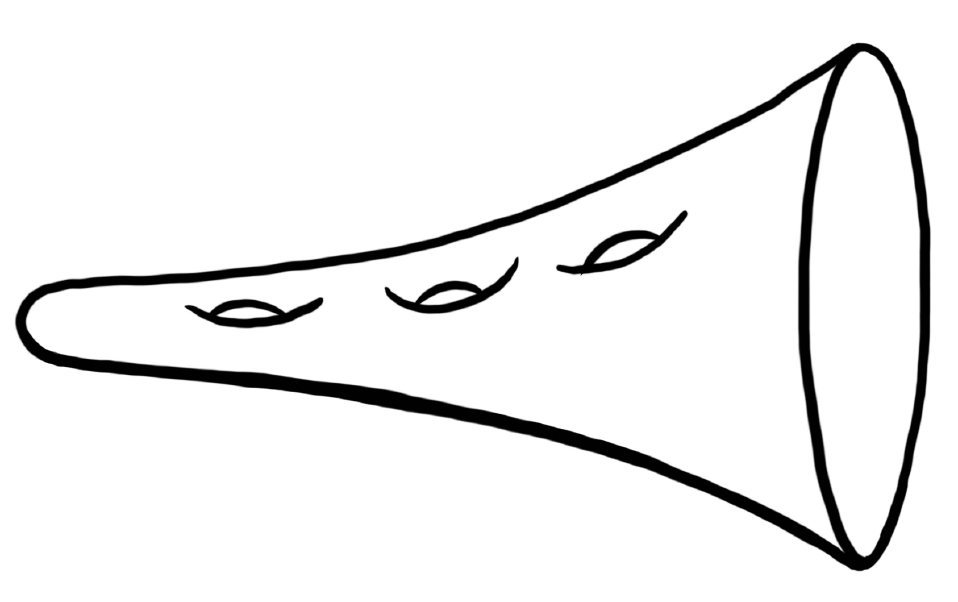}}}

\title{
On the Random Matrix Model of the Virasoro Minimal String}
\author{Clifford V. Johnson}
\email{cliffordjohnson@ucsb.edu}

\affiliation{Department of Physics, Broida Hall,   University of California, 
Santa Barbara, CA 93106, U.S.A.}


\begin{abstract}
The  model of two dimensional quantum gravity defining the ``Virasoro Minimal String", presented recently by Collier, Eberhardt, M\"{u}hlmann, and Rodriguez, was also shown to be perturbatively (in topology) equivalent to a random matrix model. 
An alternative definition  is presented here, in terms of double-scaled orthogonal polynomials, thereby allowing direct access to non-perturbative physics. Already at  leading order, the defining string equation's properties yield valuable information about the non-perturbative fate of the model, confirming that the case  $(c{=}25,{\hat c}{=}1)$ (central charges of spacelike and timelike Liouville sectors) is special, by virtue of  sharing certain key features of the ${\cal N}{=}1$ supersymmetric JT gravity string equation.  Solutions of the full string equation are  constructed using a special limit, and the (Cardy) spectral density is completed to all genus and beyond. The    distributions of the underlying discrete spectra are readily accessible too, as is the spectral form factor. Some examples of these are exhibited. 

\end{abstract}


\maketitle

\section{Introduction}

\label{sec:introduction}

There has been a considerable resurgence of interest  in the use of random matrix models to capture properties of two-dimensional quantum gravity. The  core model of interest, Jackiw-Teitelboim~\cite{Jackiw:1984je,*Teitelboim:1983ux} (JT) gravity, a theory of 2D gravity  coupled to a scalar (the ``dilaton''). The focus on it has mostly been as a gravity theory in its own right, and what might be learned about it holographically. There is also keen interest because it  arises as the near-horizon low-temperature physics of a wide class of higher dimensional black holes.   On the other hand, the double-scaled~\cite{Brezin:1990rb,*Douglas:1990ve,*Gross:1990vs,*Gross:1990aw} random matrix model techniques used by  Saad, Shenker and Stanford~\cite{Saad:2019lba}  to illuminate (perturbatvely in topology)  JT gravity have their origins in work from an earlier era, where the 2D gravity models lived on the world-sheet of various models of critical string theory. Such gravity models instead involve Liouville gravity coupled to other conformal field theories.  While it was clear that the two perspectives and approaches could be used to inform each other somewhat, many of the connections needed a clearer understanding. In particular, it has been a hope that many more of the techniques and insights gained from studies of critical string theory over the years could be brought to bear in the the dilaton gravity arena.

A new model has appeared recently that helps make considerable progress. The {\it world-sheet} description of the  ``Virasoro minimal string'' (VMS)  presented  by Collier, Eberhardt, M\"{u}hlmann, and Rodriguez~\cite{Collier:2023cyw} was shown by them to be a dilaton gravity model (of JT form but with a more general potential), giving the latter an explicit description in   critical string  terms. Moreover, the string theory can also be formulated in terms of  three dimensional chiral gravity on $\Sigma{\times}S^1$ and intersection theory theory on the moduli space of Riemann surfaces~$\Sigma$, which allows for a precise formulation of the stringy observables. They also identify a random matrix model  description of the system (perturbatively in topology), 
which generalizes that of ref.~\cite{Saad:2019lba}, but also connects nicely to the ``traditional'' role of computing correlation functions of string theory vertex operators  in terms of  intersection theory. Moreover, JT gravity itself (and its random matrix model realization) emerges in a certain classical limit  of the framework. Therefore, the  VMS  acts as a bridge between several perspectives on 2D quantum gravity.

As further motivation,  the Virasoro minimal string also has a 2D target space with some  time dependence by virtue of the presence of a {\it timelike} Liouville sector on the world-sheet. As such, it has been interpreted as a useful stringy model of cosmological spacetime (see {\it e.g.,} refs.\cite{Rodriguez:2023kkl,Rodriguez:2023wun}), making it additionally interesting as a string theory. As with the more standard $D{=}2$ string theory, having a large $N$ matrix model description is extremely useful. (See a comment on this aspect below, however.)

\begin{figure}[t]
\centering
\includegraphics[width=0.48\textwidth]{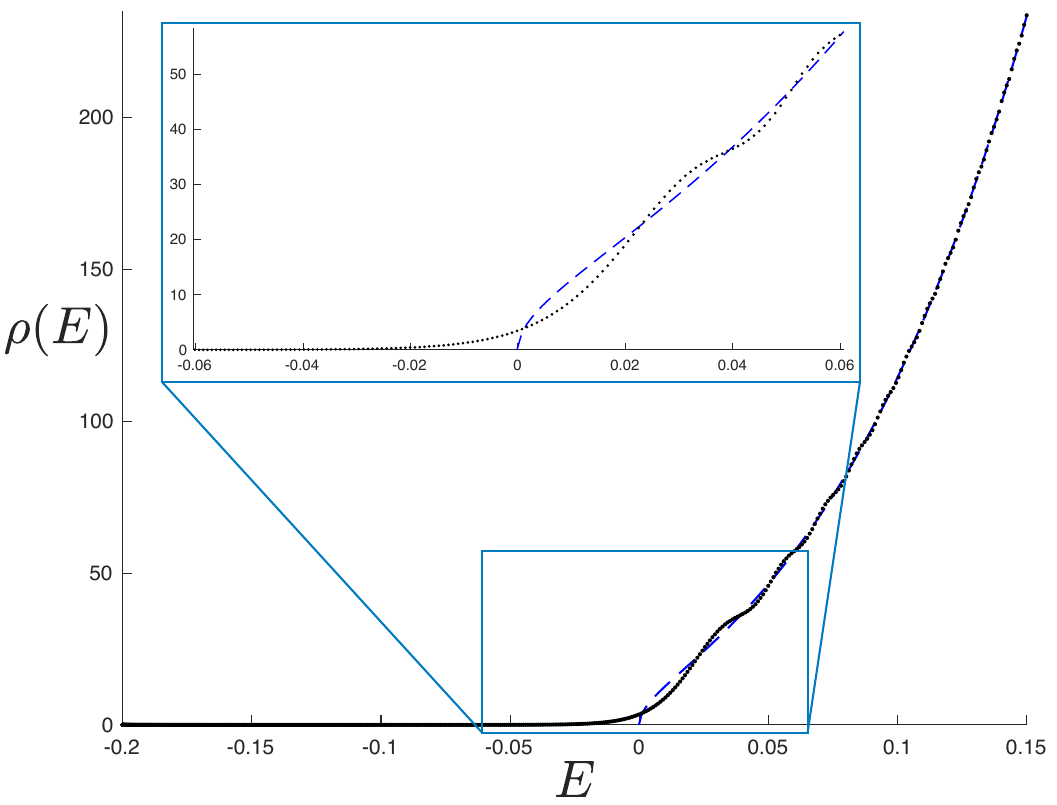}
\caption{\label{fig:full-spectral-density} The full spectral density for the $b{=}1$ Virasoro minimal string. The dashed line is the disc level result of equation~(\ref{eq:leading-spectral-density}).  Here $\hbar{=}{\rm e}^{-S_0}{=}1$. }
\end{figure}

The central focus of the present paper is to formulate the random matrix model using techniques that are not wedded to world-sheet ({\it i.e.,} topological) perturbation theory, allowing for a much wider exploration of the content of the physics. Techniques based on the (double-scaled~\cite{Brezin:1990rb,*Douglas:1990ve,*Gross:1990vs,*Gross:1990aw}) orthogonal polynomial approach, developed in refs.~\cite{Johnson:2019eik,Johnson:2020heh,Johnson:2020exp,Johnson:2021zuo,Johnson:2022wsr}  for applications in the JT gravity and supergravity context are just what are needed here. Figure~\ref{fig:full-spectral-density} gives a peek at one of the key outputs of the analysis. The case $b{=}1$ (this and other parameters will be explained below) 
is a non-perturbatively well-defined model of random Hermitian matrices  with eigenvalues on the whole real line, and  its {\it complete} spectral density  can be extracted to all orders in perturbation theory and beyond. The dashed line is the leading (disc order) spectral density~(\ref{eq:leading-spectral-density}) of the model, which is in fact the universal Cardy density of states for 2D conformal field theories. 

Many other quantities that are inaccessible in topological perturbation theory can be computed using this non-perturbative framework, and examples are explored and exhibited. These include the spectral form factor of the model, as well as the probability distributions (across the ensemble) of the individual energy levels of the  underlying microstates in the random matrix model.

It may be surprising to many that it is a {\it random matrix model} that appears here defining a 2D string theory, instead of a matrix quantum mechanics, which  is usually used for 2D strings~\cite{Brezin:1978sv, Klebanov:1991qa,Ginsparg:1993is}. These are different kinds of creature on the face of it. This is worth understanding more clearly, but  an immediate observation is that once one restricts to the singlet sector in the matrix quantum mechanics,  capturing the simple 2D string, the problem reduces to one of $N$ free fermions in an upside down quadratic potential, after double-scaling. The resulting  2D target space has the usual linear dilaton spatial direction, while the   time dimension is rather simple. On the other hand, the orthogonal polynomial treatment of the random matrix model is {\it also} equivalent to a problem of $N$ free fermions, but  in a somewhat more special multi-critical potential. It seems intuitively natural that a string theory with a 2D  target space time could emerge from such a setting too, but now with the new potential giving non-trivial behaviour in the time direction.\footnote{This could also imply that there exists a matrix quantum mechanics with non-trivial time dependence that contains the random matrix model arises as a subsector.}  This picture is worth exploring further to see if this is precisely how the timelike Liouville component arises.

\subsection{Dilaton Gravity and Liouville Gravity}
A starting point for defining the Virasoro minimal string is to consider the following two dimensional gravity plus dilaton theory~\cite{Mertens:2020hbs,Suzuki:2021zbe,Fan:2021bwt}:
\be
S=-\frac12\int_{\cal M} \sqrt{g}(\Phi {\cal R}+W(\Phi)) - \int_{\partial\cal M}\sqrt{h}\Phi(K-1) - S_0\chi\ , \nonumber
\ee with potential for $\Phi$:
\be
W(\Phi) = \frac{\sinh(2\pi b^2\Phi)}{\sin(\pi b^2)}\ ,\quad   0\leq b\leq 1\ .
\ee
Here $g$ refers to the metric on the two-dimensional (Euclidean)  spacetime $\cal M$, $h$ is the induced metric on the boundary $\partial\cal M$, and $K$ is the extrinsic curvature. The parameter $S_0$ couples to the Euler characteristic~$\chi$ of $\cal M$, so that contributions  are weighted by $\e^{S_0\chi}$. This results is a topological perturbation theory in parameter $\hbar{=}{\rm e}^{-S_0}$.  In the limit $b\to0$ the potential $W(\Phi)$ becomes linear and the theory becomes JT gravity, where the parameter~$S_0$ is the extremal entropy.

The theory has another description ({\it via } a field redefinition) in terms of two coupled  Liouville theories, one spacelike  with central charge $c{\geq} 25$, the other timelike with ${\hat c}{\leq}1$, and with Virasoro conformal weights $h_P$ and~${\hat h}_{\widehat P}$ respectively,  such that:
\bea
&&c=1+6Q^2\ , \quad {\hat c}=1-6{\hat Q}^2\ , \quad {\rm where}\nonumber\\
&&
\label{eq:bforQ}
Q=b+b^{-1}\ , \quad {\hat Q}= b^{-1}-b \nonumber\\
&&h_P=\frac{Q^2}{4}+P^2\ ,\quad {\hat h}_{\widehat P}=-\frac{{\widehat Q}^2}{4}+{\widehat P}^2\ ,
\eea 
where the Liouville momenta are related by ${\widehat P}{=}iP$,  following from the mass shell condition $h_P{+}{\hat h}_{\widehat P}{=}1$.

From this perspective, it is also natural to consider this 2D gravity theory as the worldsheet theory of a critical bosonic string theory, with $c_{\rm total}{=}c{+}{\hat c}{=}26$. In this description, the parameter $b$ now controls the exponential growth of the Liouville potentials. 

\subsection{Random Matrix Model}
Ref.\cite{Collier:2023cyw} gave a great deal of evidence that there is also a random matrix model description of this gravity theory. The model is double scaled~\cite{Brezin:1990rb,*Douglas:1990ve,*Gross:1990vs,*Gross:1990aw}, meaning that as an ensemble of  $N\times N$ Hermitian matrices $M$, the size $N$ is taken to infinity while the polynomial potential $V(M)$ is tuned to certain universal critical behaviour that yields smooth surfaces in the 't Hooftian expansion. In this description, $\hbar{=}{\rm e}^{-S_0}$ is the (renormalized) topological expansion  parameter, $1/N$. The  disc spectral density of the model is:\footnote{Ref.\cite{Collier:2023cyw} used the notation $\varrho^{(b)}(E)$, reserving the symbol $\rho$ for CFT quantity. This will not be done here.}
\be
\label{eq:leading-spectral-density}
\rho^{(b)}_{0}(E)  = \e^{S_0}{2\sqrt{2}}\frac{\sinh(2\pi b\sqrt{E})\sinh(2\pi b^{-1}\sqrt{E})}{\sqrt{E}}\ ,
\ee
which  is (with $E {=} P^2{=}h_P{-}\frac{c{-}1}{24}$),   the universal Cardy density of states for 2D CFTs.

In similar fashion to what was done by Saad {\it et. al,}~\cite{Saad:2019lba}, this spectral density serves as the seed for the full topological expansion  of matrix model quantities through a family of  recursion relations~\cite{Mirzakhani:2006fta} derivable from the general form of an Hermitian matrix model~\cite{Eynard:2007fi}. At the core of the relations are topological recursion relations for  the volumes $V^{(b)}_{(g,n)}(P_1,\ldots P_n)$ of the moduli space of certain 2D surfaces $\Sigma_{g,n}$ with $g$ handles and $n$ geodesic boundaries with lengths ${P_i}$. These ``quantum volumes'' generalize the Weil-Petersson volumes (for bordered hyperbolic Riemann surfaces) that appear in the JT case. Here they  are core observables in the string/Liouville theory setting, where the $n$ boundaries are insertions of vertex operators carrying Liouville momenta ${P_i}$, ($i=1,\ldots,n$). 

Just as for the JT gravity case, this intrinsically perturbative (in worldsheet topology) manner  of defining the matrix model, while powerful and beautiful, does not allow for much insight into physics beyond perturbation theory. This is where many questions about the physics (from the point of view of the string theory, the dilaton gravity, and also the higher dimensional black holes to which these studies connect) must be tackled.

\subsection{Anticipation and Outline} 

As already mentioned above, this paper will formulate the  matrix model a  way that does not refer to a perturbative topological structure at the outset. While it will be perturbatively equivalent to the formulation above, the framework will allow for a much wider exploration of the content of the physics.

Return to the example shown in  figure~\ref{fig:full-spectral-density} showing the {\it complete} spectral density (solid line) that can be extracted from the techniques deployed, for the case $b{=}1$ (where $c{=}25$ and $c{=}1$). In contrast to what can be done for JT gravity, this case has  a natural and {\it unambiguous} non-perturbative definition as a  model of Hermitian matrices  with eigenvalues on the whole real line. The leading disc density~(\ref{eq:leading-spectral-density}) (existing only for $E{\ge}0$) is shown as a dashed line.) There are undulations in the full non-perturbative result (the meaning of which will be explained in detail later), as well  as a tail that stretches to  all $E{<}0$.

In case reassurance is needed, note that this outcome is structurally identical to the situation  for the more well-known Airy model, the ``double-scaled'' Gaussian Hermitian matrix model, which is also non-perturbatively well-defined. It has disc density $\rho_0{\sim}\sqrt{E}$ supported only for positive $E$, while the complete spectral density lives on the whole line, with undulations, and  an exponential tail extending along the $E{<}0$ region. 

It is worth noting that  semi-classical insights into the fate of the non-perturbative definition of the model for general $b$ were  obtained in ref.~\cite{Collier:2023cyw} by analytically continuing  the disc density to $E{<}0$, beyond the classical support of $\rho_0(E)$. This gives the leading effective potential for an energy eigenvalue (now thought of as the position of a particle in a Dyson gas) in the background of all the others. (This goes back to refs.~\cite{David:1990ge,David:1991sk}, although it was revived in the JT gravity context in ref.~\cite{Saad:2019lba}). Using this, ref.~\cite{Collier:2023cyw} observed that for all cases {\it except}~$b{=}1$ there is a potential  instability to defining the ensemble on the whole real line since there are instantons representing tunneling to new minima at $E{<}0$. In the formalism used in this paper, this is nicely confirmed by examining (Section~\ref{sec:avatars}) the consistency conditions for the leading string equation using the criteria set out in refs.~\cite{Johnson:2021tnl,Johnson:2020lns}: The instability problems manifest themselves as a multi-valuedness of the leading string equation solution $u_0(x)$ in the region $x{<}0$ that renders an associated Schr\"odinger problem ill-posed. (These function and parameter $x$ will be explained below).  For $b{=}1$ there is still multi-valuedness, but it is in the so-called ``trans-Fermi'' region $x{>}0$ where it does not affect stability.

Of course, the point of this paper is that one can go far beyond semi-classical analysis and study the full string equation. The outcome is simply that there are no smooth solutions $u(x)$ to the string equation for Hermitian matrix models on the real line when $b{\neq1}$, but there is one when $b{=}1$. Constructing such a solution is difficult, but possible, and it is explained how in Section~\ref{sec:full-non-perturbative}, with results. 

It is important to note that (just as for ordinary JT gravity) this does {\it not} mean that the $b\neq1$ cases cannot be given a non-perturbative definition. The techniques of this paper suggest a very natural one as well. One simply gives up thinking of them as ensembles of Hermitian matrices on the entire real line and instead defines them with a lowest energy, denoted  $\sigma{<}0$. As long as~$\sigma$ is above the location of the point where the effective potential develops an instability, this is a perfectly well-defined non-perturbative completion of the perturbative physics and will agree with perturbation theory to all orders.  This is a natural {\it family} of completions (parameterized by $\sigma$) that differ from each other in exponentially small ways. This was originally done for JT gravity in ref.~\cite{Johnson:2019eik}, with $\sigma{=}0$, and more general $\sigma$ are explored in ref.~\cite{Johnson:2021tnl} where the framework is all explained exhaustively.

In this picture, the unambiguous ``natural'' definition of the nice case $b{=}1$  as an ensemble on the whole real line simply comes from sending $\sigma{\to}{-}\infty$. In fact, since it is easier to solve the relevant string equation for some definite value of $\sigma$, the $b{=}1$ case is solved most nicely in this way and then taking the limit.

Once the string equation is  written, defining the theory fully non-perturbatively, perturbative computations can be readily performed by solving it recursively as a non-linear ordinary differential equation. In fact, the underlying recursive structure of how solutions are developed  is explored in Section~\ref{sec:perturbation-theory} (building on refs.~\cite{Johnson:2020heh,Johnson:2021owr})  in order to show how it yields results equivalent to some of those obtained from topological recursion. Since perturbation theory is not the focus of this paper, the discussion is brief.\footnote{Alicia Castro has mentioned forthcoming perturbative work~\cite{Castro:2024kpj} on the Virasoro minimal string {\it via} a string equation, with a manuscript to appear in concert with this one. See also forthcoming work by Ashton Lowenstein on many aspects of the structure of string equations and perturbation theory, and its relation to other perturbative techniques, to appear~\cite{Lowenstein:2024}.}  Solving the string equation non-pertubatively allows a host of physical properties of the model to  be mined using various tools. The fully non-perturbative spectral density is computed in Section~\ref{sec:full-non-perturbative}, the individual probability distributions of the underlying discrete states in the ensemble are computed in Section~\ref{sec:microstates}, and the spectral form factor is computed in Section~\ref{sec:spectral-form-factor}. Some closing remarks and discussions are presented in Section~\ref{sec:closing}.

\section{Orthogonal Polynomials, Double Scaling,  and the String Equation}
\label{sec:intro-to-string-equations}

The next few paragraphs are a reminder of where some of the key elements of the double-scaled orthogonal polynomial toolbox come from, in an effort to have a self-contained narrative. Experts are welcome to skip to later parts.
The key to the approach is that random matrix models of $N\times N$ matrices $M$, with eigenvalues $\lambda_i$, can often be written in terms of  families of orthogonal polynomials~\cite{Brezin:1978sv,Bessis:1980ss},  $P_n(\lambda){=}\lambda^n+$ \{lower powers\}, characterized by a recursion relation of the form $\lambda P_n=P_{n+1}+r_nP_{n-1}$ (for even matrix model potential $V(M)$). Physical quantities in the matrix model are then expressed in terms ${\sim}N$ of the polynomials. The recursion coefficients $r_n$ defining the polynomials follow from the $V(M)$, in the form of a difference equation. In the double-scaling limit that yields gravity, $n/N$ becomes a continuous variable~$X$, and~$r_n$ becomes a function of that variable, $r(X)$. At large $N$, the eigenvalues can be treated as a (Dyson) gas of particles moving in the potential $V(\lambda)$ that repel each other due to the van der Monde determinant that arises as the Jacobian for changing variables  to the $\lambda_i$. The resulting configuration is a droplet/distribution of finite size, with endpoints $\lambda{=}\pm a$.  The potential $V(\lambda)$ can be tuned to yield certain universal critical behaviour at the endpoints, where the distribution goes as ${\sim}(\lambda \mp a)^{k-\frac12}$, where $k{=}1,2,\ldots$  
One meaning of ``universal'' here is that this form is independent of the details of the original potential $V(\lambda)$, {\it e.g.} such as whether it is quartic or cubic. This means that in the 't Hooftian interpretation of the matrix model Feynman diagrams as tessellations of the 2D spacetimes being summed over, the resulting physics in the continuum does not depend on the details, {\it e.g.} whether squares or triangles were used.

Turning to the orthogonal polynomials, the function $r(X)$ at $X{=}1$ sets the endpoint  value:  $r(X{=}1){=}r_c{\sim}a$. Scaling infinitesimally away from the end (schematically here) {\it via} $\lambda{=}{-}a{+}\delta E$, $X{=}1{-}x\delta$, $r(X) {=} r_c{-}u(x)\delta$, where $\delta{\to}0$ as $N{\to}\infty$, defines variables that describe the  universal physics near the endpoint. Here $-\infty{\leq} x{\leq}{+}\infty$. The difference  equation for $r_n$ becomes a non-linear ordinary differential equation (ODE) for $u(x)$, known 
as a ``string equation''.~\footnote{For a recent exposition in the JT gravity context, see the early sections of ref.~\cite{Johnson:2021tnl}.} 

Meanwhile, the orthogonal polynomials ($P_n(\lambda)$) themselves (times a factor of $\exp(-V(\lambda)/2)$) become  functions $\psi(E,x)$  in the limit. Remarkably, for the class of models under consideration, $\psi(E,x)$ are determined (up to  normalization) as wavefunctions of a Schr\"odinger problem: 
\be 
\label{eq:schrodinger-problem}
\left[-\hbar^2\frac{\partial^2}{\partial x^2}+u(x)\right]\psi(x,E)=E\psi(x,E)\ ,
\ee
for which $u(x)$ is the potential. The parameter $\hbar$ is the scaling piece of $1/N$ in the limit: $\delta \hbar{=}1/N$. Many things can be computed once the $\psi(E,x)$ are known. For example,  the spectral density of the model is given in terms of $\psi(E,x)$ as follows:
\be
\label{eq:spectral-density-exact}
\rho(E) =\int_{-\infty}^0 \left|\psi(x,E)\right|^2 dx\ . 
\ee
The upper limit can be some constant, denoted $\mu$, but for the purposes of this paper, $\mu{=}0$. The fact that the integral (which began life as a sum over the orthogonal polynomial index $n$) does not run over the whole range of $x$ is quite natural: While there are an infinite number of orthogonal polynomials, only a finite number ${\sim}N$ of them are needed to build the random matrix model. 

Here is a natural point for making contact with core elements of the gravity approach.
The disc order spectral density arises from taking the leading WKB form of the wavefunctions:
\bea
\label{eq:WKB-wavefunction}
\psi_0(E,x)&=& \frac{1}{\sqrt{\pi\hbar}(E-u_0)^\frac14}\times\\
&&\hskip1cm\cos\left\{\hbar^{-1}\!\!\int^{x} \!\!
\sqrt{E-u_0(x^\prime)}\,
dx^\prime-\frac{\pi}{4}\right\}\ ,\nonumber
\eea 
giving (after averaging over fast oscillations):
\be
\label{eq:spectral-density-leading}
\rho_{0}(E) \!= \frac{1}{2\pi\hbar}\int_{-\infty}^0\!\frac{\Theta(E{-}u_0(x)) dx}{\sqrt{E-u_0(x)}}=\frac{1}{2\pi\hbar}\int_{0}^E\!\frac{f(u_0)du_0}{\sqrt{E-u_0}} \ ,
\ee
where $u_0(x)$ is the leading perturbative piece of $u(x)$ obtained by sending $\hbar{\to}0$.  
In the second expression  $f(u_0){=}{-}\partial x/\partial u_0$, a Jacobian.\footnote{Notice that an $E$ integral of the leading spectral density $\rho_0(E)$ puts $\sqrt{E-u_0(x)}$ on the top line, which means that the argument of the exponential in the WKB form is precisely the leading effective potential for one eigenvalue mentioned before. This is the link between studying its properties and studying the properties of the leading string equation for $u_0$, to be uncovered below.}

Turning to the full ODE for $u(x)$ itself, the general form was derived long ago in the original double-scaling-limit papers~\cite{Brezin:1990rb,*Douglas:1990ve,*Gross:1990vs,*Gross:1990aw}, and can be written:
\be
\label{eq:little-string-equation}
{\cal R}=0\ ,\quad  {\rm where}\quad{\cal R}{\equiv}\sum_{k=1}^\infty t_k R_k[u]+x\ ,
\ee
 where the $R_k[u]$ are polynomials (see below) in the function $u(x)$ and its $x$-derivatives. 
 The $k$th model comes from a critical potential that yields the behavior $\rho_0{\sim} E^{k-\frac12}$ at the endpoint.\footnote{Put differently there is the classic $E^\frac12$ behaviour of the Gaussian case (Airy), but the coefficients in the polynomial potential can be tuned to allow $(k{s}1)$ extra zeros to land at the endpoint defining what was called ``multicritical'' behaviour in the old language.~\cite{Kazakov:1989bc}} 
 
A particular  $t_k$ determines how much the~$k$ model  contributes. These models will be treated as fundamental building blocks for constructing the matrix model potential that defines the gravity model under study. The $t_k$'s values  will be fixed shortly by matching to disc order.  

The $ R_k[u]$ are the ``Gel'fand-Dikii''~\cite{Gelfand:1975rn} differential polynomials  in $u(x)$ and its derivatives, normalized here so that the non-derivative part has unit coefficient: $R_k{=}u^k+\cdots+\#u^{(2k-2)}$ where $u^{(m)}$ means the $m$th $x$-derivative. For example:
\bea
\label{eq:gelfand-dikii-R}
&&R_1{=}u \ ,\quad  R_2{=}u^2{-}\frac{\hbar^2}{3}u^{\prime\prime}\ ,\nonumber\\
&&R_3 {=} u^3+\frac{\hbar^2}{2}(u^\prime)^2+{\hbar^2}uu^{\prime\prime}+\frac{\hbar^4}{10}u^{(4)}\ , \cdots
\eea
 Successive $R_k[u]$ can be obtained using a recursion relation, which will be useful later:
 \be
 \label{eq:gelfand-dikii-recursion}
 R^\prime_{k+1} = \left(\frac{2k+2}{2k+1}\right)\left[\frac12u^\prime R_k+u R^\prime_k-\frac{\hbar^2}{4}R_k^{\prime\prime\prime}\right]\ .
 \ee

\section{Determining the  String Equation}
\label{sec:the-string-equation}
\subsection{The Leading String Equation}
\label{sec:leading-string-equation}

The next steps proceed in the familiar manner done for  earlier matrix model studies of ordinary JT and various ${\cal N}{=}1$ JT models\cite{Johnson:2019eik,Johnson:2020mwi,Johnson:2020exp}. 
Disc perturbation theory is the regime $x{\to}{-}\infty$ and the  leading string equation for $u_0(x)$ is simply 
\bea 
\label{eq:tree-string-eq}
{\cal R}_0\equiv\sum_k t_k u_0^k+x=0\ ,
\eea 
in that regime.
 So the $t_k$ are uniquely determined by requiring that the leading spectral density coming from the integral~(\ref{eq:spectral-density-leading}) that uses $u_0(x)$ as input yields  equation~(\ref{eq:leading-spectral-density}). One way to proceed is to  simply expand $\rho_{0}(E)$ as a power series in $E^\frac12$, and use the fact that for the $k$th critical model, the string equation in the negative~$x$ region is $u_0^k{+}x{=}0$, so $f(u_0){=}ku_0^{k-1}$, and  integral~(\ref{eq:spectral-density-leading}) gives~\cite{Johnson:2020heh} $\rho_0^{(k)}(E){=}C_kE^{k-\frac12}/2\pi\hbar$, where $C_k{=}2^{2k{-}1}k ((k{-}1)!)^2/(2k{-}1)!$. Matching coefficients for the full expansion gives:
\be
\label{eq:new-teekay}
t_k=2\sqrt{2}\pi\frac{\pi^{2k}}{(k!)^2}
\left(Q^{2k}-{\widehat Q}^{2k}\right)\ ,
\ee
where $Q$ and ${\widehat Q}$ are given in terms of the gravity/Liouville parameter $b$ in equation~(\ref{eq:bforQ}).

\begin{figure}[t]
\centering
\includegraphics[width=0.48\textwidth]{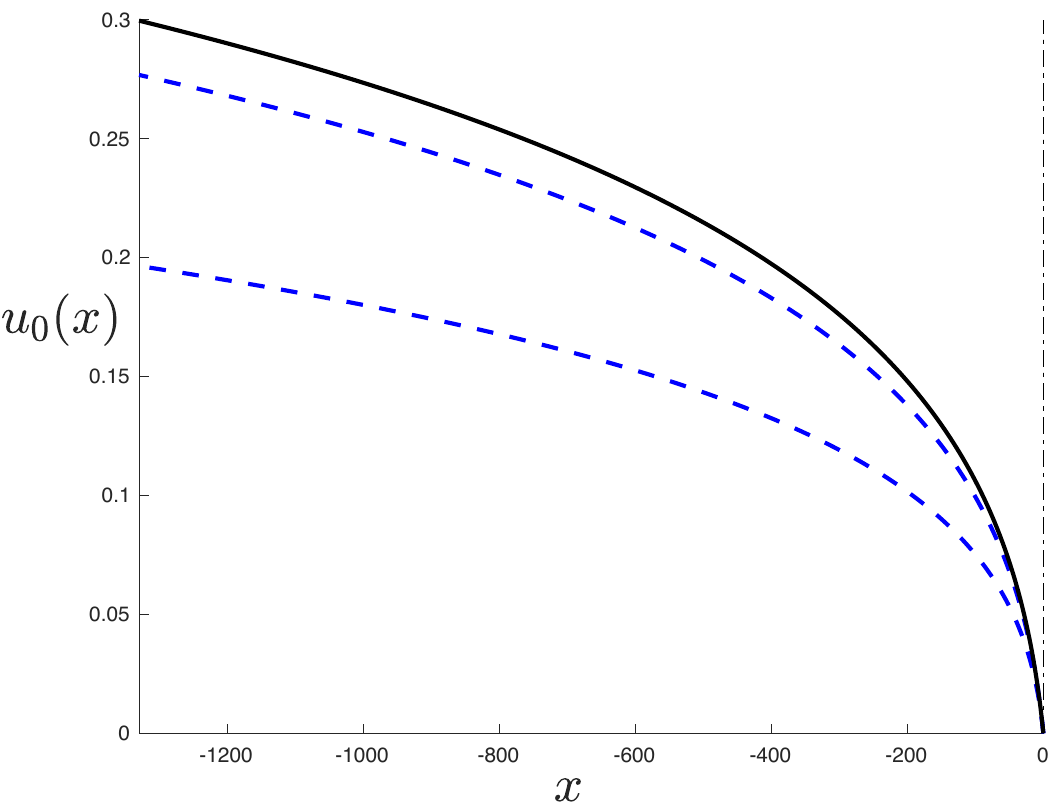}
\caption{\label{fig:leading-u-plots} The leading potential  $u_0(x)$ for (lowest) the $b{=}\frac12$ case, $b=0.75$, and (highest) the $b{=}1$ case.}
\end{figure}

Rather nicely, the equation for the $t_k$ can be used to write this new equation for $u_0(x)$ in  closed form:~\footnote{Another, more direct way to get this is to use an integral transform that acts on $\rho_0(E)$ to give $f(u_0)$, an ``inverse Abel transform". It may be found in ref.~\cite{Johnson:2020lns}, along with a derivation.}
\bea
2\sqrt{2}\pi\left[I_0(2\pi Q\sqrt{u_0}\,)-I_0(2\pi{\widehat Q}\sqrt{u_0}\,)\right]+x=0\ .
\label{eq:leading_string_equation}
\eea
where $I_0(y)$ is  the modified Bessel  function in $y$ of order~$0$. Some examples for positive $u_0$ are shown in figure~\ref{fig:leading-u-plots}.

A key remark here is that when $b{\to}1$, $Q{\to}2$ and ${\hat Q}{\to}0$ and then the form of the  $t_k$ and of the tree level string equation is the same (up to a trivial rescaling) as that discovered in refs.~\cite{Johnson:2020heh,Johnson:2020exp} for ${\cal N}{=}1$ JT supergravity:
\bea
2\sqrt{2}\pi\left[I_0(4\pi \sqrt{u_0})-1\right]+x=0\ .
\label{eq:leading_string_equation_super}
\eea
This will have some nice consequences shortly. It is important to note that this does {\it not}
mean that the model actually is secretly supergravity in this limit. The matrix model in that case is not of the $\beta{=}2$ Dyson class, as this is. Crucially, for the supersymmetric case the integrals~(\ref{eq:spectral-density-exact},\ref{eq:spectral-density-leading}) run some finite amount into the $x>0$ region, which is crucial for defining the hard edge  $1/\sqrt{E}$  leading behaviour appropriate to such models. Instead, for this paper the upper limit is zero, appropriate to a bosonic, $\beta{=}2$, Dyson ensemble.

\subsection{The Complete String Equation}
For double-scaled Hermitian matrix models (with eigenvalues on the real line), the form of the string equation as an ODE was already given in equation~(\ref{eq:little-string-equation}), but for general $t_k$. Now that the $t_k$ have been determined as expression~(\ref{eq:new-teekay}) by the leading  string equation by comparing to the disc order spectral density, the full equation is now fixed. Whether there are solutions beyond perturbation theory or not is a matter to be returned to in Sections~\ref{sec:avatars} and~\ref{sec:full-non-perturbative}.

\section{Avatars of Non-Perturbative Physics}
\label{sec:avatars}

In diagnosing potential issues of instability, there is a useful connection   between the effective potential analysis of the random matrix model on the one hand, and properties of the solution $u_0(x)$ of the leading string equation~(\ref{eq:leading_string_equation}).  The $u_0(x)$ of the Virasoro minimal string turns out to be an excellent case  to study, as a function of parameter $b$.

First, consider the possibility that at some value of~$x$, there are multiple values of $u_0(x)$. A crucial observation made in ref.~\cite{Johnson:2021tnl} (expanding on remarks in ref.~\cite{Johnson:2020lns}) is that it is inconsistent to have the function $u_0(x)$ be multi-valued in the regime $x{<}0$. In such a case, the Schr\"odinger problem~(\ref{eq:schrodinger-problem}), where $u_0(x)$ is a potential, does not make sense.   Furthermore,  the presence of multi-valuedness also translates into a problem of finding smooth  solutions to the full string equation that connect to small $\hbar$ perturbation theory---if  such a solution existed, the limit $\hbar{\to}0$ cannot unambiguously connect such a smooth solution to a multivalued one.

On the other hand, multivaluedness in the regime $x{>}0$ (the ``trans-Fermi regime''\footnote{So called since it refers to orthogonal polynomial indices well beyond the top ($N$th) one, which defines a Fermi level in an alternative presentation of the physics where the orthogonal polynomials are Slater determinant realizations of many body fermion wavefunctions~\cite{Banks:1990df}.}) is less of a concern since that region is non-perturbative compared to the $x{<}0$ regime, and so properties of the string equation that turn on beyond perturbation theory can significantly modify the solution there.

\begin{figure}[t]
\centering
\includegraphics[width=0.48\textwidth]{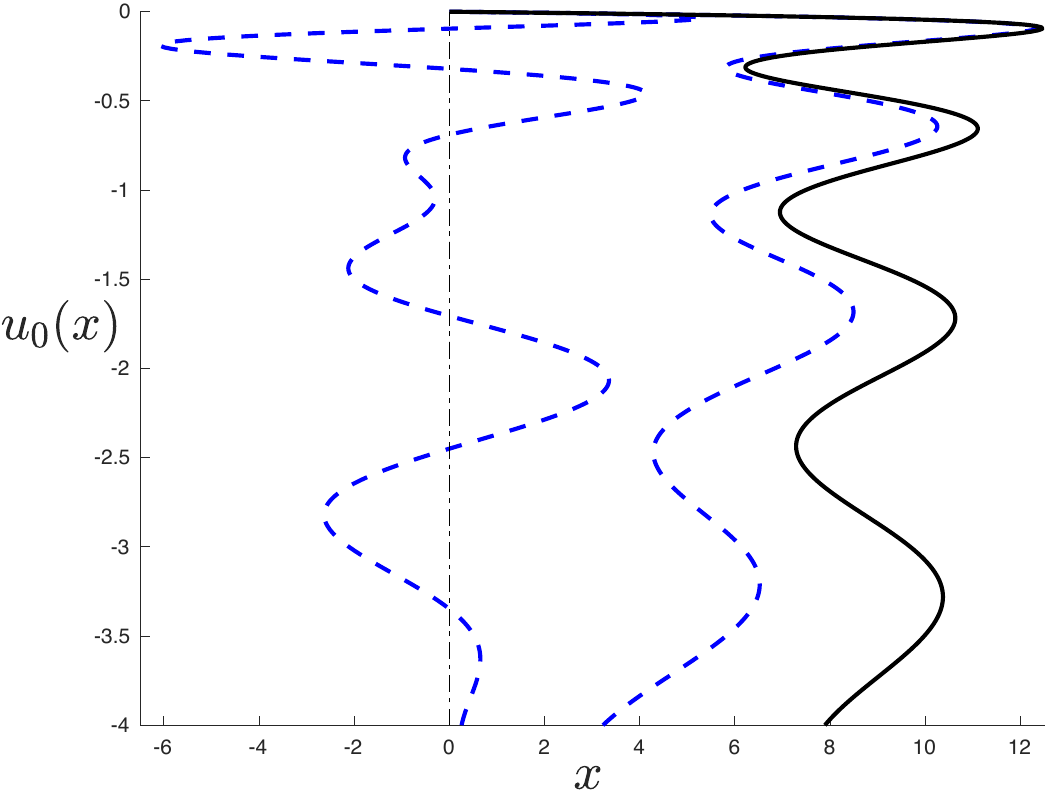}
\caption{\label{fig:comparing-u-wiggles} The leading potential  $u_0(x)$ for (leftmost) the $b{=}\frac12$ case, $b=0.94$, and (rightmost) the $b{=}1$ case, showing the multivaluedness that develops after $u_0(x)$ becomes negative. After starting undulations at the origin, all curves except the $b{=}1$ case  will eventually cross the $x{=}0$ line again.}
\end{figure}

The Virasoro minimal string provides a very clear illustration of all this. Simply put, for all cases of $b{<}1$, the function $u_0(x)$ defined by the tree level string equation~(\ref{eq:leading_string_equation}) develops undulations that generate multi-valuedness for $u_0(x)$ in the $x{<}0$ regime. This follows from the fact that after a sign change in $u_0(x)$, because of the square root in the defining equation~(\ref{eq:leading_string_equation}), the modified Bessel function becomes an ordinary Bessel function, with its accompanying  oscillatory behaviour. This is already known for the case $b{=}0$ (since that is  JT gravity). Rewriting the leading string equation in the negative $u_0$ regime, one has the family of curves parameterized by $b$:
\bea
x=2\sqrt{2}\pi\left[J_0(2\pi {\widehat Q} \sqrt{-u_0}\,)-J_0(2\pi Q\sqrt{-u_0}\,)\right]\ ,
\label{eq:leading_string_equation-under}
\eea
where as a reminder $Q{=}b^{-1}{+}b$ and ${\widehat Q}{=}b^{-1}{-}b$. Some examples are shown in figure~\ref{fig:comparing-u-wiggles}. The key observation  is that although they all pass through the origin heading toward positive $x$, the piece controlled by ${\widehat Q}$ will always eventually pull $x$ back below zero.  This takes longer to happen as $b{\to}1$, but at $b{=}1$ the ${\widehat Q}$ component vanishes and then $x{=}2\sqrt{2}\pi(1-J_0(4\pi\sqrt{-u_0})){\geq}0$ always since $J_0$ only approaches 1 when its argument is zero. So in this special case  the multi-valuedness is entirely contained in the $x{>}0$ regime, where it is not harmful. (Since the $b{=}1$ case is of the same form as the ${\cal N}{=}1$ JT supergravity leading string equation, its ``safe'' multivaluedness has already been observed~\cite{Johnson:2021tnl}.)

The conclusion is that there can be no solutions to the full non-perturbative string equation~(\ref{eq:little-string-equation}) for the cases $b{<}1$ and hence the random matrix model of the  Virasoro minimal string is (as already noticed in ref.~\cite{Collier:2023cyw} and recalled in the Introduction) ill-defined as a model of Hermitian matrices with eigenvalues {\it on the whole real line}.

It is crucial to note that this does {\it not} mean that the $b{<}1$ models cannot be given a non-perturbative definition, just not as Hermitian matrices with eigenvalues along the whole real line. This will be returned to later.

\section{Perturbative Expansions}
\label{sec:perturbation-theory}
Before going on to uncover fully non-perturbative physics, some remarks about recovering perturbation theory are in order. Once the leading solution $u_0(x)$ to the string equation is known, perturbation theory  with content equivalent to that obtained using topological recursion in ref.~\cite{Collier:2023cyw} can be developed by using the full string equation~(\ref{eq:little-string-equation}). The fact that it is built from Gel'fand-Dikii polynomials $R_k[u]$ will be useful in a moment.

The function $u(x)$ is simply the second $x$--derivative of the closed string partition function~${\widehat Z}$:
\be
\label{eq:closed-partfun}
u(x)=\hbar^2\frac{\partial^2{\widehat Z}}{\partial x^2}\ ,
\ee
A ``closed string'' topological perturbative expansion for $u(x)$ implies one for ${\widehat{Z}}$:
\bea 
\label{eq:topological-expansion}
u(x) &=& u_0(x)+\sum_{g=1}^\infty u_{2g}(x)\hbar^{2g}+\cdots\ ,\nonumber\\
\widehat{Z}&=&
\sum_{g=0}^\infty \widehat{Z}_{g}\hbar^{2g-2}+\cdots\ , 
\eea
(the ellipses denote non-perturbative contributions)
with $\widehat{Z}_{\rm sphere}{=}\widehat{Z}_0\hbar^{-2}$,  $\widehat{Z}_{\rm torus}{=}\widehat{Z}_1$, and so on, where $g$ is the number of handles on the surface. 

The partition function of Euclidean 2D gravity, however, is the Laplace transform of the spectral density:
\be
\label{eq:part-fun}
Z(\beta){=}\int \rho(E){\rm e}^{-\beta E} dE\ ,
\ee
It is the expectation value of ${\rm Tr}({\rm e}^{-\beta M})$ where $M$ is the matrix. It is an {\it asymptotic}
loop operator with  finite-size boundary of length $\beta$. 
Since  it is an infinite sum of powers of $M$, it can also be thought of as having turned on a specific combination of an infinite family of point-like operators,~\footnote{This is the microscopic-macroscopic loop connection introduced in ref.~\cite{Banks:1990df}.} denoted ${\cal O}_k$ in the ``KdV'' basis, meaning that as a function of the $t_k$ now thought of as operator coefficients, the generalized KdV flows for $u(x;\{t_k\})$ are:
 \be
 \label{eq:kdv-flows}
 \frac{\partial u}{\partial t_k}\simeq \frac{\partial R_{k+1}[u]}{\partial x}\ .
 \ee
 In individual critical models, $t_k$ derivatives correspond to insertions of the point-like operators ${\cal O}_k$ on the worldsheet (2D spacetime~$\Sigma$), and the PDE above represents the RG flow between critical models. Here, and in other dilaton gravity theories captured in the way done here, something subtly different must be going on, since all the $t_k$ (and hence all the ${\cal O}_k$) are turned on in a specific combination. This suggests that there must be another kind of finite loop present in the model. It is natural to point to the geodesic boundaries that are present in the decomposition into asymptotic boundaries connected by trumpets to bordered Riemann surfaces. In the Virasoro minimal string they are of lengths $\{P_i\}$ corresponding to the amount of Liouville momentum contributing to that part of the correlator. In JT gravity they were denoted~$\{b_i\}$ in ref.~\cite{Saad:2019lba} (not to be confused with the Liouville parameter $b$ in use here).

Operationally, perturbation theory works as follows. The topological expansion~(\ref{eq:topological-expansion}) can be inserted into the string equation~\ref{eq:little-string-equation}, and using the fact that $u_0(x)$ is known from matching to the spectral density (see equation~(\ref{eq:leading_string_equation})), the $u_{2g}(x)$ can be computed iteratively. (Some of this was explored in this way in the ${\cal N}{=}1$ JT supergravity context in ref.~\cite{Johnson:2020lns}.) 
For example, beyond $u_0(x)$, the next (torus) contribution $u_2(x)$ can be computed by substituting into the string equation, and expanding, keeping all terms at order $\hbar^2$. A useful  result  for the Gel'fand-Dikii polynomials at this order can be derived~(see {\it e.g.,} ref.\cite{Johnson:2020lns}) by using the recursion relation they satisfy and it is:
\bea
\label{eq:Rk-expansion}
&&R_k[u] = u^k-\frac{\hbar^2}{12}k(k-1)u^{k-3}\left[ 2uu^{\prime\prime}+(k-2)(u^\prime)^2\right] \nonumber\\
&&\hskip 5.5cm +O(\hbar^4)\ .
\eea
and after substitution, 
the terms at order $\hbar^0$ vanish since $u_0(x)$ satisfies the equation at that order, and then requiring the $\hbar^2$ order to vanish gives an equation for $u_2(x)$ in terms of $u_0(x)$ and its  $x$ derivatives:
\bea
&&u_2\sum_k t_k k u_0^{k-1}= \frac{1}{12}\left[\sum_{k}2t_k(k-1) u_0^{k-2}u_0^{\prime\prime}\right.\\
&&\hskip2.5cm +
\left.\sum_k t_k k (k-1)(k-2)u_0^{k-3}(u_0^\prime)^2\right]\ , \nonumber
\eea
The sums over $k$ can in fact be written as  $u_0$-derivatives of the non-trivial function of $u_0$ in the tree-level string equation~(\ref{eq:tree-string-eq}), {\it i.e.,} the part involving Bessel functions in equation~(\ref{eq:leading_string_equation}). Writing that equation as $G(u_0){=}{-}x$, and using a dot to denote a derivative with respect to $u_0$, the relation is now:
\bea
\label{eq:second-order}
u_2{=}\frac{1}{12}\left[2\frac{\ddot G}{\dot G}u_0^{\prime\prime}
+\frac{\dddot G}{\dot G}(u_0^\prime)^2\right]
{=}\frac{1}{12}\left[-2\frac{(\ddot G)^2}{(\dot G)^4}+
\frac{\dddot G}{({\dot G})^3}\right]
\eea
where $x$-derivatives of $u_0$ were traded for $u_0$ derivatives of $G(u_0)$
in the last line. Using relation~(\ref{eq:closed-partfun}) the torus contribution can therefore be written:
\be
\label{eq:torus}
{\hat Z}_{\rm torus} = \frac{1}{12}\ln {\dot G}(u_0)\ =-\frac{1}{12}\ln u_0^\prime,
\ee
(where this succinct form has been noticed in much earlier literature, {\it e.g.} ref.~\cite{Belavin:2010pj}).
Expanding ${\dot G}(u_0){=}{-}1/u_0^\prime$ in the case in hand (recall $G$ is the LHS of equation~(\ref{eq:leading_string_equation})) gives the leading $u_0$ dependence: 
\bea
\label{eq:leading-G}
{\dot G}(u_0)&=& 2\sqrt{2}\pi^2\left(1+\frac{\pi^2}{2} \left(b^2+\frac{1}{b^2}\right)u_0+\cdots\right)\nonumber \\
&=& 2\sqrt{2}\pi^2\left(1+2\pi^2 \frac{(c-13)}{24} u_0+\cdots\right)
\ ,
\eea
which after discarding a constant and expanding the logarithm, is the expected form of the $b$ dependence for the torus. With a suitable normalization, and after introducing a $P_1$ insertion (a geodesic boundary) this should match with the ``quantum volume'' expression in ref.~\cite{Collier:2023cyw} for the insertion of momentum on the torus,  $V^{(b)}_{1,1}(P_1)=\frac{1}{24}\left(\frac{(c-13)}{24}+{P_1^2}\right)$. 

One way to explore that is by using another useful tool in this formalism, the Gel'fand--Dikii resolvent ${\widehat{R}}(x,E)$, which is another direct route to the spectral density.
It satisfies the following equation~\cite{Gelfand:1975rn}
\be
4(u-E)\widehat{R}^2-
2\hbar^2\widehat{R}\widehat{R}^{\prime\prime}+\hbar^2(\widehat{R}^\prime)^2=1\ ,
\ee
and
\be
\label{eq:resolvent-density}
\rho(E) = \frac{1}{\pi\hbar}{\rm Im}\int_{-\infty}^0\!\!\widehat{R}(x,E)dx\ .
\ee
\begin{widetext}
Solving for $\widehat{R}(x,E){=}\sum_{g=0}^\infty \widehat{R}_g(x,E)\hbar^{2g}+\cdots$ recursively for a given $u(x)$ yields, after choosing a sign for the root:\footnote{This corrects some typographical omissions of derivative factors in v1 of this manuscript.}
\bea 
{\widehat{R}}(x,E) = -\frac12\frac{1}{[u(x)-E]^{1/2}}-\frac{\hbar^2 }{64}\left[\frac{5(u(x)^\prime)^2}{[u(x)-E]^{7/2}}-\frac{4u(x)^{\prime\prime}}{[u(x)-E]^{5/2}} \right]
-\frac{ \hbar^4}{4096} \left[ \frac{1155(u(x)^\prime)^4}{[u(x)-E]^{13/2}} +\cdots
\right]+\cdots \nonumber
\eea
where (since they won't be needed here) terms involving higher derivatives of $u(x)$ have been omitted from the order $\hbar^4$ terms.
\end{widetext}
As a warm-up to  see how this works, the case $u(x){=}{-}x$ is the classic Airy model.  The integrals in equation~(\ref{eq:resolvent-density}) are elementary (the lower limit should be cut off at $x{=}{-}E$) and the result is
\be
\label{eq:airy-expanssion}
\rho_{\rm Ai}(E) = \frac{\sqrt{E}}{\pi\hbar}+\frac{1}{32\pi}\frac{\hbar}{E^{5/2}}+\frac{105}{2048\pi}\frac{\hbar^3}{E^{11/2}}\cdots
\ee
which is the result of expanding the well-known exact result for the spectral density:
\be
\rho_{\rm Ai}(E)=\hbar^{-2/3}
\left(
{\rm Ai^\prime(-\zeta)^2-\zeta{\rm Ai}(-\zeta)^2}\right)\ ,\,\, \zeta=-\hbar^{-2/3}E
\ .
\ee The first term in expansion~(\ref{eq:airy-expanssion}) is the contribution from one boundary (disc), while the second is the ``torus'' with one boundary, and so forth. 

It is amusing to connect this to the topological recursion formalism, where (using the notation of ref.~\cite{Saad:2019lba}) writing $E{=}{-}z_1^2$ the relevant objects are the $W_{g,n}$, and relevant here are:
\be
W_{0,1}=2z_1^2\ ,\,W_{1,1}=\frac{1}{16z_1^4}\ ,\,\,{\rm and}\,\, W_{2,1}=\frac{105}{1024 z_1^{10}}\ ,
\ee
which are related to the resolvents $R_{g,1}$ defined there as $W_{g,1}{=}{-}2z_1R_{g,1}(-z_1^2)$.
So a comparison shows that $R_{g,n}(E){=}{-}\int^0_{-\infty} \widehat{R}_{g}(x,E)dx$. 

The $W_{g,1}$ are Laplace transforms: $W_{g,1}(z_1){=}\int_0^\infty b_1 {\rm e}^{-b_1 z_1}V_{g,1}(b_1)$, defining the (formal in this case) volumes that are analogues of the Weil-Petersson quantities for the JT case:
\be
V_{1,1}=\frac{b_1^2}{96}\ , \,\,{\rm and}\,\,V_{2,1}=\frac{105}{1024}\frac{b_1^8} {9!}\ ,
\ee
($V_{0,1}$ is undefined).

It should now be  clear that the Gel'fand-Dikii equation, starting with $u(x)$ for the Virasoro minimal string, will yield $\omega^{(b)}_{g,1}(z_1)$, the Laplace transforms of the quantum volumes $V^{(b)}_{g,1}(P_1)$ of ref.~\cite{Collier:2023cyw}. Now there are additional $\hbar$ corrections from the expansion of $u(x){=}u_0(x){+}u_2(x)\hbar^2+\cdots$, so taking that into account, to order $\hbar^2$ (to get the torus with boundary):
\begin{widetext}
\bea 
\label{eq:gelfand-dikii-A}
{\widehat{R}}(x,E) = -\frac12\frac{1}{[u_0(x)-E]^{1/2}}
+
\frac{\hbar^2}{64}\left\{
\frac{16u_2(x)}{[u_0(x)-E]^{3/2}}
-\frac{5(u^{\prime}_0(x))^2}{[u_0(x)-E]^{7/2}}+\frac{4u^{\prime\prime}_0(x)}{[u_0(x)-E]^{5/2}}\right\}
+\cdots\ ,
\eea
\end{widetext}
where $u_0(x)$ solves the tree level string equation~(\ref{eq:leading_string_equation}), and $u_2(x)$ was found earlier in terms of $u_0(x)$ (see equation~(\ref{eq:torus})). The next step is to integrate with respect to~$x$. The first term of course gives $\rho_0^{(b)}(-E)$ by construction (this was the work of Section~\ref{sec:leading-string-equation}), and hence yields (in the notation of ref.~\cite{Collier:2023cyw}) $\omega^{(b)}_{0,1}$, after multiplying by ${-}2z_1$. The remaining terms should yield, up to factors, ref.~\cite{Collier:2023cyw}'s $\omega^{(b)}_{1,1}(z_1)$ after multiplying by ${-}2z_1$:
\be
\label{eq:omega-one-one}
\omega^{(b)}_{1,1}(z_1)=\frac{1}{(2\pi)^4}\left(  
\frac{c-13}{48 z_1^2}+\frac{3}{(4\pi)^2}\frac{1}{z_1^4}
\right)\ ,
\ee and happily this does indeed work, as will be seen in what follows. In preparation, it is worth noting that something special must happen at higher genus: The expected results are simply finite polynomials in inverse powers of $z_1{=}(-E)^\frac12$.  As before, the $x$ integrals will get  their  finite contribution from the $x{=}0$ limit, and these will involve $u_0(x)$ and its derivatives evaluated there, yielding the polynomials' coefficients. It is therefore prudent to expand the leading string equation~(\ref{eq:leading_string_equation}) around there for later use:
\bea
\label{eq:useful-expansion}
 u_0(x)=-\frac{\sqrt{2}}{4\pi^2}x-\frac{1}{16\pi^2}\left(\frac{c-13}{6}\right)x^2\cdots 
\eea where the $b$ dependence was converted into the dependence on the central charge $c$. 

Note that the expansion~(\ref{eq:gelfand-dikii-A}) can be thought of as a rewriting of the classic resolvent expansion~\cite{Gelfand:1975rn} of Gel'fand and Dikii, which is:
\be
\label{eq:gelfand-dikii-B}
{\widehat{R}}(x,E) =\sum_{k=0}^\infty N_k\frac{R_k[u(x)]}{(-E)^{k+\frac12}}\ ,\quad N_k\equiv\frac{(2k-1)!}{(-4)^kk!(k-1)!}\ ,
\ee
where $N_k$ translates the normalization for the differential polynomials $R_k$ used here to that used in ref.~\cite{Gelfand:1975rn}. Pulling out a factor $(-E)^\frac{n}{2}$ in each denominator of equation~(\ref{eq:gelfand-dikii-A})  and then expanding the remainder as a power series in $u(x)/E$ reconstructs equation~(\ref{eq:gelfand-dikii-B}).
Since the results for the objects $\omega_{g,1}(z_1)$ are simply inverse powers of $(-E)^\frac12$, it makes sense to work with  expansion~(\ref{eq:gelfand-dikii-B}). Recursion properties of the $R_k[u(x)]$ will the turn out to be useful for pulling out the pieces needed to construct the $\omega^{(b)}_{g,1}$ at each order. The price paid however, is that the  expansion as it stands is not ordered nicely in $\hbar$. Every one of the infinite terms contributes a piece at a given $\hbar$ order, re-summing to give expansion~(\ref{eq:gelfand-dikii-A}). This is easily overcome~\cite{Johnson:2021owr} by temporarily doing a Laplace transform from $E$ to $\beta$:
\bea
\label{eq:efficient}
{\cal L}({\widehat R}(x,E)) = &&\!\!
\frac{1}{2\sqrt{\pi}\beta^{\frac12}}\sum_{k=1}^\infty\frac{(-\beta)^k}{k!}R_k[u(x)]\ ,\\
&&\hskip-2.5cm= 
\frac{{\rm e}^{-\beta u(x)}}{2\sqrt{\pi}\beta^{\frac12}}\left[1-\frac{\hbar^2}{12}\left(2\beta^2u(x)^{\prime\prime}-\beta^3(u(x)^\prime)^2\right)+\cdots\right]\ ,\nonumber
\eea
where in the last step the $\hbar$ expansion of $R_k[u(x)]$ in equation~(\ref{eq:Rk-expansion}) was used, and then a re-summation performed.
Now perturbation theory can be studied rather efficiently. For example, writing $u(x){=}u_0(x){+}\hbar^2u_2(x)+\cdots$ gives at order $\hbar^2$ a term with the following pieces:
\be
{\rm e}^{-\beta u_0(x)}\left( 12\beta u_2(x)+2\beta^2u_0(x)^{\prime\prime}-\beta^3(u_0(x)^\prime)^2\right)\ ,
\ee
but since   $u_2(x){=}{-}\frac{1}{12}\frac{d}{dx}(u_0^{\prime\prime}/u_0^\prime)$, (from results~(\ref{eq:second-order}) and (\ref{eq:torus})) this term is in fact a total $x$-derivative of ${\rm e}^{-\beta u_0(x)}\left( -\beta u_0(x)^{\prime\prime}/u_0(x)^\prime+\beta^2u_0(x)^{\prime}\right)$. After integrating and transforming from $\beta$ back to $E$ and restoring overall factors, the genus one contribution is finally:
\be
\int_{-\infty}^0\!\!R_{1}(x,E)dx=\hbar^2
\left[
\frac{1}{32} \frac{u_0^\prime(0)}{(-E)^{5/2}}
-\frac{1}{48} \frac{u_0^{\prime\prime}(0)/u_0^\prime(0)}{(-E)^{3/2}}
\right]\ ,
\ee
which after multiplying by ${-}2z_1{=}{-}2(-E)^{1/2}$, and using the expansion~(\ref{eq:useful-expansion}) gives the desired result~(\ref{eq:omega-one-one}) for $\omega^{(b)}_{1,1}(z_1)$ (up to an overall numerical factor attributable to the conventions  of ref.~\cite{Collier:2023cyw}.) A Laplace transform  (from~$z_1$ to $P_1$)  turns this into ref.~\cite{Collier:2023cyw}'s ``quantum volume" $V^{(b)}_{1,1}(P_1)$, completing the demonstration. 

While the computation to this order could have been done more swiftly directly using expansion~(\ref{eq:gelfand-dikii-A}), going by this route illustrates two key features that {\it must} generalize to all orders, showing that this approach to the matrix model's perturbation theory is equivalent to what can be obtained from topological recursion techniques. The first is the organization in terms of inverse powers of~$z_1$, automatically built into   the prototype expansion~(\ref{eq:gelfand-dikii-B}) of Gel'fand and Dikii, and the second is the recursion relation~(\ref{eq:gelfand-dikii-recursion}) among the differential polynomials $R_k[u(x)]$. It was responsible for the reorganization of perturbation theory into the efficient form given in equation~(\ref{eq:efficient}), {\it and} the relation between the correction  $u_2(x)$ and $u_0(x)$ and its derivatives. These two pieces worked together to produce the total derivative, giving the coefficients of the powers of $z_1$ in terms of $u_0$ and its derivatives at $x{=}0$. This is a non-trivial result and the expectation that it  persists to higher orders (in order to be equivalent to topological recursion results) is likely guaranteed by the underlying Gel'fand-Dikii recursion. It is of value to pursue more aspects  of this correspondence.

It is interesting to explore  other perturbative properties of the Virasoro minimal string using this formalism, but the discussion of the full non-perturbative physics, unlocked by the formulation of this paper in particular,  should be delayed no further.

\section{Full Non-Perturbative Physics}
\label{sec:full-non-perturbative}

The next step is to find a solution to the string equation~(\ref{eq:big-string-equation}) that connects smoothly to the perturbative physics of the $x{<}0$ regime. This is to be understood as the process of finding the complete recursion function from which the entire set of orthogonal polynomials can be determined. Such a solution should be well-defined on the whole real $x$ line, so that the Schr\"odinger problem~(\ref{eq:schrodinger-problem}) can yield a full set of wavefunctions. Such a solution should  exist for $b{=}1$, but what will be developed here will have relevance to the $b{<}1$ cases as well, in a manner that will be clarified later.

The string equation is formally of infinite order, since each $t_k$ controls a term with $2k{-}2$ derivatives and all the~$t_k$ are turned on ($k{=}1,\ldots,\infty$). As discussed in ref.~\cite{Johnson:2020exp}, however, since the $t_k$ decrease in size as $k$ increases, a sensible truncation of the equation can be done that can capture the physics up to any desired accuracy. In solving the equation, it is natural to take the boundary condition on the $x{<}0$ side to asymptotically be the leading string equation (truncated at some high order), but the boundary condition on the $x{>}0$ side is subtle, since due to the oscillatory behaviour in the intermediate regime (for negative $u_0$), a truncation to $k$ odd is quite different from one to $k$ even. Odd $k$ produces a well defined boundary value problem for the equation since the curve settles down to a tail that asymptotes to positive large~$x$, while even $k$ does not. Even so, it is not clear how under control the truncation is in this $x{>}0$ regime.

A different approach affords more control. Instead, one can study the string equation for the ensemble of Hermitian matrices with a lowest eigenvalue, call it $\sigma$, and have $\sigma{<}0$, as well as ensuring that the solution still has ${\cal R}{\to}0$ in the $x{\to}{-}\infty$ region, which {\it makes sure  to recover the same  perturbation theory}. This has the beauty of providing a non-perturbative definition for all the $b$, as long as (for a given $b$) the $\sigma$ is above the first well of the effective potential. 

For the case of $b{=}1$, the limit $\sigma{\to}{-}\infty$ can be taken with impunity, providing a non-perturbative definition as a random matrix model of Hermitian matrices with eigenvalues on the whole real line.

The string equation required is:\footnote{ This equation, with $\sigma{=}0$, first arose in early studies~\cite{Dalley:1991qg,*Dalley:1992br,*Dalley:1991vr,*Morris:1991cq,*Dalley:1991xx,*Anderson:1991ku} of {\it positive} matrix model ensembles for gravity applications. It was used to provide a non-perturbative defintion of JT as an ensemble with lowest energy 0. Non-zero $\sigma$ was later begun to be understood in ref.\cite{Dalley:1992yi}, and solutions first thoroughly explored in ref.\cite{Johnson:1992wr}. The perspective that it simply defines Hermitian matrix models with lowest eigenvalue $\sigma$ allowed for an exploration of a family of non-perturbative definitions of JT gravity in \cite{Johnson:2021tnl}. This same equation (with $\sigma{=}0$ and a term $\hbar^2\Gamma^2$ on the right hand side) can also be used to define random matrix ensembles of class  $(1{+}2\Gamma,2)$ in the AZ classification), but crucially, the integrals that extract the physics, such as in equation~(\ref{fig:full-spectral-density}) run to some finite positive~$x$, not zero. This  allowed for the study of  ${\cal N}{=}1$ and ${\cal N}{=}2$  JT supergravity in refs.~\cite{Johnson:2020heh,Johnson:2020exp,Johnson:2023ofr}.}
\be
\label{eq:big-string-equation}
(u-\sigma){\cal R}^2-\frac{\hbar^2}2{\cal R}{\cal R}^{\prime\prime}+\frac{\hbar^2}4({\cal R}^\prime)^2=0\ .
\ee 
Where ${\cal R}$ is still defined as in equation (\ref{eq:little-string-equation}), with the $t_k$ chosen as in equation~(\ref{eq:new-teekay}) for  study of the Virasoro minimal string. 
Notice that ${\cal R}{=}0$ for all $x$ is a solution to this equation, but this is not the solution to use.  The relevant boundary conditions on $u(x)$ are:
\bea
&&{\cal R}[u]\to{\cal R}_0[u_0] = 0\ , \quad {\rm as}\quad x\to-\infty\nonumber \\
&&u(x) = \sigma\ , \quad {\rm as}\quad x\to+\infty
\eea
As $|\sigma|$ gets larger, notice that the term which dominates most in the equation is  $\sigma{\cal R}^2$, ensuring that in the limit of large negative $\sigma$, the $u(x)$ solution indeed becomes that of ${\cal R}{=}0$, defining an ensemble of random hermitian matrices with eigenvalues on the whole real line.

\begin{figure}[t]
\centering
\includegraphics[width=0.48\textwidth]{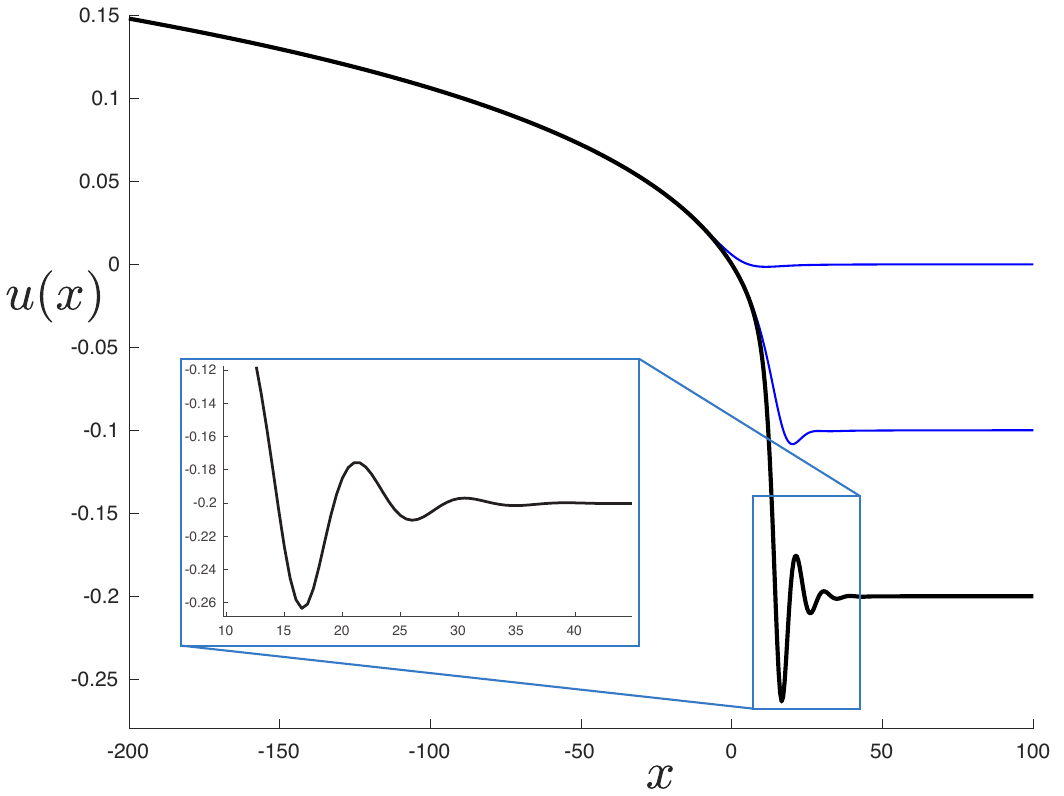}
\caption{\label{fig:full-potential} The potential  $u(x)$ for the $b{=}1$ case, computed for $\sigma{=}0,{-}0.1,{-}0.2$ (upper,middle, lower). The inset shows more details of the well in the intermediate regime for $\sigma{=}{-}0.2$.   Here $\hbar{=}{\rm e}^{-S_0}{=}1$. }
\end{figure}

With so many words said, it is time to show the results. With a truncation to $k{=}7$, one can readily (enough) solve the (14th order) equation for $u(x)$, and the case of $b{=}1$ with $\sigma{=}0,{-}0.1,{-}0.2$ are shown in figure~\ref{fig:full-potential}, with the choice $\hbar=1$.\footnote{Other cases of $b$ were also solved, but the curves are not significantly different from each other to warrant displaying them.}  

Some features are worth noting. (1) The undulations near the well of the solution are {\it not} the result of numerical instability: The solution, obtained \footnote{The {\tt bvp4c} routine within {\tt MATLAB} was used, with vectorization turned on, and with Jacobians input by hand to aid with stability. Solving took ${\sim}156$ to ${\sim}{565}$ seconds, depending upon $\sigma$. For more details on handling this equation, see refs.\cite{Johnson:2020exp,Johnson:2022wsr}} on an $x$ grid of $1.3{\times}10^5$ points for $-3000\leq x\leq +3000$, is accurate to within a reported absolute tolerance of ${\sim}5{\times}10^{-7}$. (2) The  value of $\sigma{=}{-}0.2$ is already large and negative enough to see that the contribution to the density at this point is deep into the exponential tail: The only parts of the wavefunctions that will contribute to the $x<0$ region at these energies are small exponentially suppressed tunnelling contributions, since at $x=0$, $u(x){\simeq}0$, and rising fast for smaller $x$. 

The Schr\"odinger problem~(\ref{eq:schrodinger-problem}) was solved using $u(x)$ as potential, obtaining a set of wavefunctions $\psi(E,x)$ for 800 energy points. After normalization (by matching to the known analytic form in the large positive $x$ regime), the spectral density was computed by doing integral~(\ref{eq:spectral-density-exact}), with the result  given in figure~\ref{fig:full-spectral-density}. Going to smaller $\sigma$ is possible, extending the tail further to the left, and with the solution for $\rho(E)$ for the parts already computed making ever-so-slight adjustments that are smaller than the resolution of the currently drawn curves. In short, the limit as $\sigma{\to}{-}\infty$ is fully under control for the $b{=}1$ case, as already discussed above. 

For $b{<}1$ cases, decreasing $\sigma$ past a certain point can begin to produce a new peak corresponding to the results of the suggested eigenvalue tunneling seen in the semi-classical analysis. Analogues of this have been fully understood and analyzed in the recent work presented in ref.~\cite{Johnson:2022pou} and so will not be explored again here. The upshot is that for suitably chosen  $\sigma\leq0$, the $b<1$ cases have a perfectly fine non-perturbative definition using this same methodology as ensembles of Hermitian matrices with lowest eigenvalue $\sigma$. 

For the rest of the paper, the special case of $b=1$ will be the focus in specific examples, but it should be borne in mind that most statements and computations have application to any value of $b$, with a suitable choice of $\sigma$ for supplying a non-perturbative defintion.

\section{Microstates from Fredholm Determinants}
\label{sec:microstates}

A much more general object that is computable from the current definition of the random matrix model is the Kernel:
\be
\label{eq:kernel}
K(E,E^\prime) = \int_{-\infty}^0\!\!  dx \, \psi(E,x)\psi(E^\prime,x) \ .
\ee
It contains much more information than the spectral density, which is merely its diagonal. Its derivation and uses are reviewed in this context in ref.~\cite{Johnson:2022wsr}.
In a discrete model before double-scaling,  $K$ is a finite  matrix, and  probabilities of the locations of eigenvalues can be phrased as determinants of $K$ or submatrices of it. After double scaling $K$ plays the same role, now getting promoted to an infinite dimensional operator acting on functions on the $E$ line according to  
$
\int_a^b K(E,E^\prime)f(E^\prime) dE^\prime{=} g(E)
$,
on some interval $(a,b)$. Denoting  $\mathbf{K}_{(a,b)}$ as the integral operator with kernel $K(E,E^\prime)$ on the interval, it is a classic result~\cite{Gaudin1961SurLL} that the ``gap'' probability that there are no energies in the interval is given by the Fredholm determinant ${\rm det}(\mathbf{I}-\mathbf{K}_{(a,b)})$. 
 \begin{figure}[t]
\centering
\includegraphics[width=0.48\textwidth]{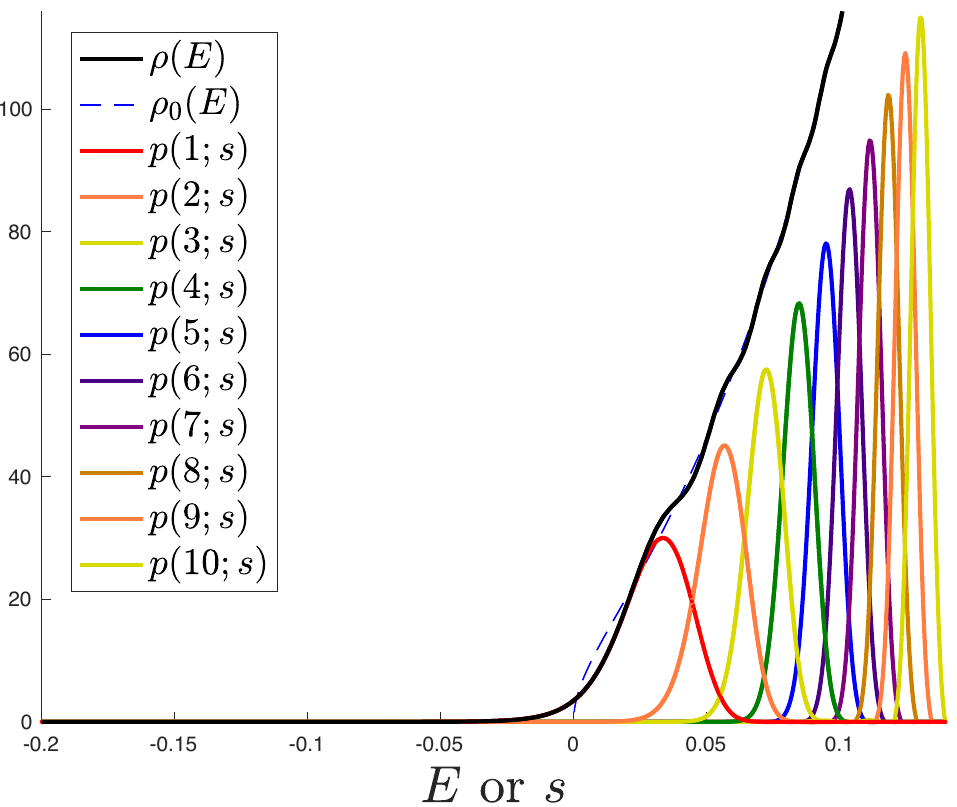}
\caption{\label{fig:microstate-peaks} The  first ten probability distribution peaks $p(n,s)$, $n{=}1,2,\ldots,10$ for individual energy levels across the ensemble described by the random matrix model, computed using Fredholm determinants. The black solid line is the sum of the peaks, the spectral density $\rho(E)$. The dashed line is $\rho_0(E)$. This is for the $b{=}1$ case. Here $\hbar{=}{\rm e}^{-S_0}{=}1$. }
\end{figure}
Choosing $a{=}{-}\infty$ and $b{=}s$, a reference energy, and writing the Fredholm determinant in this case as $E(1;s){=} {\rm det}(\mathbf{I}-\mathbf{K}|_{(-\infty,s)})$, this is the cumulative probability distribution for the lowest (first) energy of the ensemble. The probability density  function  for finding  an energy is thus  $p(1;s){=}{-}dE(1;s)/ds.$ A famous example of this is the Tracy-Widom distribution for the Airy model~\cite{Tracy:1992rf}. In fact, the probability distribution for the $n$th energy level can be computed iteratively from these tools, and will be denoted $p(n,s)$.

This techniques were first used for the study of gravity in refs.~\cite{Johnson:2021zuo,Johnson:2021rsh}, where the underlying microstate distributions for JT gravity and various JT supergravity were uncovered. This fully non-perturbative (from the point of view of the gravity or string topological expansion) exercise allows for a re-appreciation of the spectral density as a discrete sum of peaks:
\be
\rho(E) = \sum_{n=1}^\infty p(n,E)\ ,
\ee
giving the understanding that the non-perturbative bumps/undulations seen in the spectral density are simply these microstate peaks added together.

The same can be done here for the Virasoro minimal string,  since the wavefunctions $\psi(E,x)$ have now been computed, and so $K(E,E^\prime)$ can be computed. Methods for carefully computing the Fredholm determinant numerically in this setting are reviewed in ref.~\cite{Johnson:2022wsr} (the work of ref.~\cite{Bornemann_2009} is particularly helpful), and the first ten microstate peaks for the Virasoro minimal string for the case $b{=}1$ are displayed   in figure~\ref{fig:microstate-peaks}. Indeed, the bumps in the exact spectral density align precisely with the peaks, as they should.

An interesting additional novel aspect of this is the fact that since the Virasoro minimal string  spectral density is also the universal Cardy distribution for a conformal field theory, the random matrix model  provides data on the underlying microscopic distributions that underlie that distribution, which should have meaning in  its own right for the study of conformal field theory. It would be interesting to explore this further.

 \section{The Spectral Form Factor}
 \label{sec:spectral-form-factor}
 
Another quantity of interest that can be readily computed using the non-perturbative formulation is the  spectral form factor:
\bea
\label{eq:spectral-form-factor}
&&{\rm SFF}(\beta,t)\equiv\langle Z(\beta{-}it)Z(\beta{+}it)\rangle\\
&&= \langle Z(\beta{-}it)Z(\beta{+}it)\rangle_{\rm d/c.}+
\langle Z(\beta{-}it)Z(\beta{+}it)\rangle_{\rm c.}\ , \nonumber
\eea 
a sum of disconnected (d/c.) and connected (c.) pieces, where 
$
Z(\beta)
$
is the 2D gravity partition function~(\ref{eq:part-fun}), which in the present formalism is given by:
\be
Z(\beta)=\langle {\rm Tr}\,\,{\rm e}^{-\beta{ M}}\rangle=\int_{-\infty}^0 \langle x|{\rm e}^{-\beta {\cal H}}  |x\rangle
={\widehat{\rm Tr}}(e^{-\beta{\cal H}}{\cal P})\ ,
\ee
where ${\cal H}$ is the Schr\"odinger operator in equation~(\ref{eq:schrodinger-problem}), and the 
${\widehat{\rm Tr}}$ is in the $|x\rangle$ basis. The shorthand  ${\cal P}{\equiv}\int_{-\infty}^0 dx\, |x\rangle \langle x| $.
The two-point loop correlator in terms of which the spectral form factor is defined  has a disconnected piece and a connected piece, as indicated in equation~(\ref{eq:spectral-form-factor}). At leading order, diagrammatically,  the former is two copies of the disc while the latter is the cylinder. Up to a Laplace transform, the fully non-perturbative partition function has already been discussed above, so two copies go into the disconnected piece. The non-perturbative connected part of the two-point function is given in general as~\cite{Banks:1990df}:
\begin{eqnarray}
\label{eq:correlator-connected}
&\langle&\!\!\!\!\! Z(\beta) Z(\beta^\prime)\rangle_{\rm c.} = {\widehat{\rm Tr}}(e^{-\beta{\cal H}}(1-{\cal P}) e^{-\beta^\prime{\cal H}}{\cal P}) \\
&=&\!\!\! {\widehat{\rm Tr}}(e^{-(\beta+\beta^\prime){\cal H}}) - {\widehat{\rm Tr}}(e^{-\beta{\cal H}}{\cal P} e^{-\beta^\prime{\cal H}}{\cal P})\nonumber \\ 
&=& \!\!\! Z(\beta{+}\beta^\prime) -\!\! \int \!\!dE\!\!\int \!\!dE^\prime {\rm e}^{-\beta E}K(E,E^\prime) {\rm e}^{-\beta^\prime E^\prime} K\!(E^\prime,E)
\ ,\nonumber
\end{eqnarray}
where the Kernel $K(E,E^\prime)$ was defined in equation~(\ref{eq:kernel}). 
Hence, substituting $\beta\to\beta{+}it$ and $\beta^\prime\to\beta{-}it$,
\bea
&&\langle Z(\beta+it) Z(\beta-it)\rangle_{\rm c.}=  \\
&&Z(2\beta)-\!\! \int \!\!dE\!\!\int \!\!dE^\prime {\rm e}^{-\beta(E+ E^\prime)}{\rm e}^{-it(E- E^\prime)}|K(E,E^\prime)|^2 \nonumber
\ ,
\eea
which at late times will become $Z(2\beta)$, the  plateau value.

\begin{figure}[t]
\centering
\includegraphics[width=0.48\textwidth]{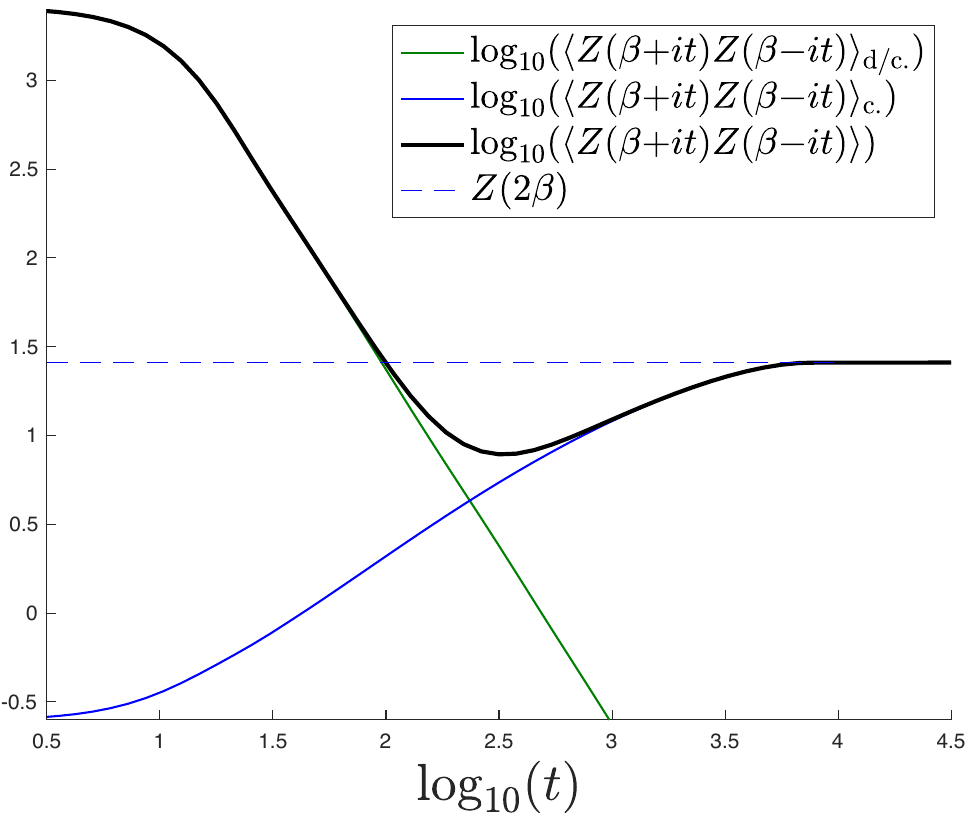}
\caption{\label{fig:spectral-form-factor} The   log of the spectral form factor for the $b{=}1$ case, with $\beta{=}10/3$. The falling green curve is the (log of the) disconnected part, while the rising blue represents the connected part. The dashed line is the value of the plateau at late time, $Z(2\beta)$. Here $\hbar{=}{\rm e}^{-S_0}{=}1$. }
\end{figure}

From this, using the wavefunctions $\psi(E,x)$ computed numerically earlier, the spectral form factor can be computed. For the $b{=}1$ case, the result is shown in figure~\ref{fig:spectral-form-factor}.  The disconnected part dominates at early times, while the connected piece is subdominant. At a transition time they exchange dominance, and the total  rises until saturation at the expected  plateau, representing the universal average behaviour of ultra-low energy two-point correlation of the underlying ensemble of matrix model spectra at late times. While the turnover to the plateau is generically non-perturbative, the leading part of the rising (ramp) part is computed in gravity (or string theory) as the universal wormhole/cylinder diagram, just as in the prototype case in JT gravity~\cite{Saad:2019lba}.

\section{Closing Remarks}
\label{sec:closing}
This paper has presented a  fully non-perturbative definition of the double-scaled random matrix model of the Virasoro minimal string of ref.~\cite{Collier:2023cyw}. The methods, based on orthogonal polynomials,  also constitute an alternative toolbox for computing many perturbative results. For the case  $b{=}1$, the random matrix model is a $\beta{=}2$ (Dyson) ensemble of Hermitian 
matrices with eigenvalues on the whole real line. The non-perturbative string equation that defines it is of the classic  Painlev\'{e}~I hierarchy ``multicritical'' form first derived in refs.~\cite{Brezin:1990rb,*Douglas:1990ve,*Gross:1990vs,*Gross:1990aw}. Notably, for $b{=}1$ the  precise admixure of critical models (set by~$t_k$) allows for smooth well-behaved solutions, even with the presence of $k$ even cases which individually fail to do so.

For the cases $b{<}1$, the ensemble can be non-perturbatively defined using the same methods, but on the line $[\sigma,+\infty)$ where $\sigma{\leq}0$ is an adjustable non-perturbative parameter of the definition. (Different  choices give exponentially small differences in the non-perturbative physics.) JT gravity itself, with  previously presented non-perturbative completions of this kind~\cite{Johnson:2019eik,Johnson:2021tnl}, is the case $b{=}0$ here.

As mentioned in the Introduction, the Virasoro minimal string is an important bridge between various approaches to two dimensional quantum gravity, connecting models of critical string theory world-sheets (and the associated techniques to studying them) to models of dilaton gravity. Regarding double-scaled random matrix models as continuum limits of dynamical tessellations of 2D Euclidean surfaces, it was natural for them to appear in both approaches, but it was nevertheless surprising for them to appear in so familiar a form when first used for JT gravity in ref.~\cite{Saad:2019lba}. In a sense, the Virasoro minimal string and the fact that it can be captured by a random matrix model helps to more clearly demonstrate that random matrix models are a much more generally applicable and  powerful tool than is widely appreciated in the quantum gravity  (and string theory) literature. This is particularly apparent if one adopts the non-perturbative point of view (emphasized in ref.~\cite{Johnson:2022hrj}) that they are an effective tool in the spirit of Wigner, going beyond the perturbative 't Hooftian view that is more directly connected to tessellations of 2D surfaces. 

It is to be expected that random matrix models will find precise roles in more settings, in various dimensions, and this is especially  where coarse-graining over a discrete spectrum is happening. For example, the Virasoro minimal string random matrix model's spectral density is the universal Cardy density of states in a 2D CFT. In retrospect, it could have been arrived at by simply taking the Wignerian approach and deriving a random matrix model with that coarse-grained spectrum by reverse-engineering the required potential, as was done in this paper. This approach (which seems to have broader applicability and flexibility than ref.~\cite{Saad:2019lba}'s method of taking limits of specific minimal models, although it was inspired by that method) can likely be more generally applied.

Finally, it is also notable that the Virasoro minimal sting is a 2D string theory, with what appears to be a time dependent target space. A special case of it ($b{=}1$) has been interpreted~\cite{Rodriguez:2023kkl,Rodriguez:2023wun} as a cosmology, although this would seem to be applicable to all $b$.
As discussed in the Introduction, exploring the details of how the 2D target spacetime description emerges from a random matrix model (instead of a matrix quantum mechanics) could help shed more light on the nature of the spacetime, and what might be learned from  its description as a tractable  string theory background.

\begin{acknowledgments}
CVJ  thanks   the  US Department of Energy for support (under award \#\protect{DE-SC} 0011687), Ashton Lowenstein and Felipe Rosso for comments, and  Amelia for her support and patience.    
\end{acknowledgments}

\bibliographystyle{apsrev4-1}
\bibliography{Fredholm_super_JT_gravity1,Fredholm_super_JT_gravity2}

\end{document}